# Remodeling of Fibrous Extracellular Matrices by Contractile Cells: Predictions from Discrete Fiber Network Simulations

Abhilash Nair[1], Brendon M. Baker[2], Britta Trappmann[2], Christopher S. Chen[2] and Vivek B. Shenoy[1*]

[1]Department of Materials Science and Engineering, University of Pennsylvania, Philadelphia, PA 19104, USA, [2]Department of Biomedical Engineering, Boston University, Boston, MA 02215, USA

**Abstract**

Contractile forces exerted on the surrounding extracellular matrix (ECM) lead to the alignment and stretching of constituent fibers within the vicinity of cells. As a consequence, the matrix reorganizes to form thick bundles of aligned fibers that enable force transmission over distances larger than the size of the cells. Contractile force-mediated remodeling of ECM fibers has bearing on a number of physiologic and pathophysiologic phenomena. In this work, we present a computational model to capture cell-mediated remodeling within fibrous matrices using finite element based discrete fiber network simulations. The model is shown to accurately capture collagen alignment, heterogeneous deformations, and long-range force transmission observed experimentally. The zone of mechanical influence surrounding a single contractile cell and the interaction between two cells are predicted from the strain-induced alignment of fibers. Through parametric studies, the effect of cell contractility and cell shape anisotropy on matrix remodeling and force transmission are quantified and summarized in a phase diagram. For highly contractile and elongated cells, we find a sensing distance that is ten times the cell size, in agreement with experimental observations.

1. Introduction

Interaction of cells with ECM is fundamental to many physiological processes such as cancer metastasis, fibrosis and wound healing [1]. These interactions are the basis of the concept of dynamic reciprocity during which cells deform and reorganize the matrix, while matrix remodeling feeds back to modulate cell contractility. The ECM is both hierarchical and fibrous with each level of hierarchy possessing characteristic constitutive and failure responses. Major challenges in accurately modeling the mechanical behavior of the ECM and their interactions with cells are the non-affine nature of the matrix deformations, remodeling of the matrix in response to cellular forces and the ability of the cells to sense the surroundings and modulate the level of the contractile forces they exert. Deformation of cells depends on the dynamic interplay between cell-related factors (e.g., cell shape, contractility, and signaling) and extracellular factors (e.g., chemical and mechanical properties of the ECM). For biomedical applications and design of novel materials, it is important to understand the mechanisms behind the observed response and to develop mathematical and computational models that can faithfully capture the positive feedback between cell contractility and matrix remodeling.

---

[*] Corresponding Author: vshenoy@seas.upenn.edu

One of the striking experimental observations in fibrous ECMs such as collagen and fibrin gels is the transmission of forces over relatively long distances [2]. When cells embedded in collagen matrices contract, the displacement fields are felt as far as 20 times the cell radius [3], [4]. Fibers aligned with the principal direction of loading undergo axial stretching while transverse fibers bend or in extreme cases, buckle. As a result, the response is highly non-affine and deformations propagate longer distances along the principal direction. These aligned bundles bear most of the forces and this leads to further alignment of the connected fibers as the deformation progresses. The ensuing response is highly non-linear and heterogeneous, as the aligned bundles are much stiffer compared to the average nominal matrix stiffness and such inhomogeneity makes the response of the materials different from conventional synthetic non-fibrous materials such as hydrogels. This mechanical response has relevance to cell functions such as migration, differentiation, and proliferation in both healthy as well as disease conditions[5]–[7]. For example, in the progression of a carcinoma, transformed epithelial cells proliferate uncontrollably, eventually breaching through the basement membrane upon which they encounter the fibrillar ECM of the collagen-rich stroma. While it is has been often observed that organized collagen fibrils within the cancerous microenvironment appear to encourage directed tumor cell migration [8]–[10], an understanding of how the collagen is reorganized and how the alignment of collagen fibrils within the tumor stroma guides tumor cell migration does not exist.

In the case of a matrix with multiple cells, within a critical distance, cells start to communicate with each other through the aligned fiber bundles and then further deformation takes place along the aligned tracts of collagen fibers that form between the cells [5]. In collagen matrices, mammary acini interact over long distances through the formation of aligned bundles of collagen fibers/ collagen tracts. The transition of the phenotype of acini to invasive type and its disorganization depends on such interaction between the neighboring acini. The shapes of the contracting cells are influenced by the mechanical feedback and their disease state. Invasive and non-invasive cancer cells adopt different shapes and induce anisotropic strain energy density distribution in the surrounding collagen matrices during migration and invasion. Invasive cells are long and spindle shaped with highly anisotropic distribution of strain energy aligned with the long axis of the cell [11]. In the case of non- invasive cells, the shapes tend to be spherical with more isotropic strain energy distributions.

There have been some recent studies about the role of cell shape and contractility in the matrix reorganization and force transmission [12]–[14]. However, most of these models do not capture the process of matrix reorganization; our goal here is to model this process and to develop a computational framework that bridges the gap between experiments and idealized theoretical models. Sanders deduced the theoretical bounds of the deformation zone around a spherical contracting cell[12]. His calculation is applicable in the limits of linear elasticity and symmetric cells with uniform contraction. The contribution of fibrous matrices to the elastic response has been considered in the work of Ma et al. by including the distribution of fibers obtained from experiments into finite element models [3]. Gjorevski and Nelson assigned heterogeneous stiffness to material points based on experimental measurements[15]. With these approaches, their models are able to qualitatively show the range of force transmission in collagen gels. However, these simulations do not predict how a network that is initially random reorients and realigns to applied loads. Barocas and coworkers have elicited the role of the matrix reorganization and fiber alignment using a multi-scale modeling approach [13], [16]–[18]. Through their work on cell compacted gels, Aghvami *et al.*, qualitatively showed the experimentally observed fiber alignment patterns in gel seeded with multiple cells [13]. The heterogeneity in the gel during off- axis and equibiaxial stretch have been shown by Sanders *et al.* using the same multi-scale model [16]. Multi-scale models account for the fibers in the matrix in a coarse-grained average sense. The fibrous networks in these models are used to obtain constitutive response for each material point based on the macroscopic deformation of the networks; the models do not track the deformation of individual fibers. This aspect can be limiting if the buckling wavelengths of the fibers are larger than the mesh size of the coarse-grained model. As we show below, fiber buckling is a commonly observed feature in the case of anisotropically contracting cells. In these cases it behooves us to track fibers individually. Furthermore, no models to date have predicted how the range of force transmission depends on the mechanical properties of the matrix and the contractility and shape of the cells.

Our objective in this work is to develop a model that naturally accounts for the process of matrix reorganization resulting from cell-mediated fiber alignment and stretching, and use this model to predict the deformation zone around a contractile cell. The typical length scales of collagen gels enable a direct modeling instead of a coarse-grained approach. Our model shows the formation of experimentally observed collagen alignment and bundling from initially random fiber distributions, thereby enabling long-range force transmission. Using this modeling approach, we systematically study the role of cell shape anisotropy and cell contractility on the zone of influence surrounding the cells. From the insights gained from the simulations of the response of individual cells, we extend the model to multicellular systems and predict the critical contraction required for cell-cell interactions to occur as a function of cell and matrix properties.

**2. Computational Model**

The response of fibers and networks in 2D and 3D have been studied for homogenous loads such as simple shear and uniaxial extension by Onck et al [19] and Hu et al [20] respectively. The qualitative responses, in particular, non-affine elastic deformations at small strains followed by affine stretching of fibers with increasing loads, is similar in both dimensions. However, 2D offers more confinement than 3D leading to a higher critical buckling load for fibers and a response dominated less by bending in the case of networks. Thus, in addition to possessing the advantage of being computationally simpler, 2D models capture all aspects of network mechanics including non-affine stiffening, fiber alignment and bending-stretching transitions. Therefore, for a systematic exploration of the large parameter space of cell and matrix properties in cell-populated matrices, we have developed a finite element (FE) based 2D discrete fiber network (DFN) model. Following our earlier work on active biopolymer networks [21], [22] the 2D fiber networks representing collagen gels are created with randomly organized linear elastic fibers and rigid crosslinks. Fibers of length *L* are uniformly distributed in the computational domain of size *W* much larger than the fiber length (*W/L* >16) and a crosslink is formed when two fibers intersect[24]. Collagen fibers have diameters in the range of few 100 nanometers to few microns and moduli of few 100 MPa [16], [25], [26]. As the persistence length of collagen fibers is in the range of few microns, these fibers are typically modeled as linearly elastic fibers. Fibers are modeled using shear flexible Timoshenko beam elements in the finite element package ABAQUS [27]. Each section of the fiber between two crosslinks is discretized into 4 elements (determined based on the convergence studies). Total strain energy during deformation is given by

$$U = \frac{1}{2} \sum_{i=0}^{fibers} \int \left[ EI \left(\frac{\partial \psi_i(s)}{\partial s}\right)^2 + EA \left(\frac{\partial u_i(s)}{\partial s}\right)^2 + \lambda GA \left(\frac{\partial v_i(s)}{\partial s} - \psi_i(s)\right)^2 \right] ds$$

where, $E$ is the young's modulus, $G$ is the shear modulus and $I$ is the second moment of area of the fibers, $\partial u/\partial s$ is the axial strain, $\partial v/\partial s$ is the rotation of the fiber cross-section, $\psi(s)$ is the rotation of the plane perpendicular to the normal axis of the fiber, $\lambda$ is the shear correction factor, $u$ is the axial displacement along the filament axis, $s$ and $v$ is the displacement along the two transverse axes.

In order to model isotropic cell contraction, a region corresponding to the cell is removed from the network and displacement boundary conditions are applied at the tips of the fibers intersecting the cell boundary. The representative concentrations of the collagen gel used in the simulation are 1-5 mg/ml and a parametric study is carried out by varying the aspect ratios of the cells ($a/b$) from 1 to 16 (see Appendix A for the conversion of collagen concentration into equivalent network density and response of gels with various concentrations). This corresponds to changing the cell shape from circular to a highly polarized spindle shape and in all cases, the "volume" of the cell, $\pi ab$ is assumed to be constant. The outer edges of the network are rigidly fixed as shown in Figure 1. In all simulations, boundary effects are eliminated by ensuring that the fibers at the boundary of the computational domain experience negligible strains. For a given cell size, multiple simulations are carried out to ensure that the results obtained are independent of the domain size. This required that with increasing cell aspect ratio, the domain size be correspondingly increased to eliminate boundary effects. Figure 3 shows the cell aspect ratios and the computational domain size used in the simulations.

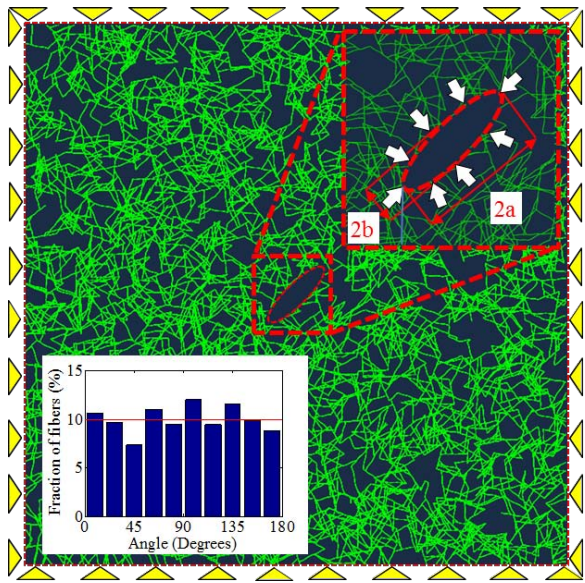

Figure 1: Collagen gel is modeled using 2D networks of elastic fibers with random orientations (histogram of the fiber distribution is given in the left inset) with edges rigidly fixed. The region corresponding to the elliptical cell is removed and displacement boundary conditions are applied to simulate isotropic cell contraction. The right inset shows the cell shape and boundary conditions.

## 3. Results

To understand the deformation modes of the networks due to contracting cells we first consider the case of a uniformly contracting elliptical cell. Here, cellular forces lead to large deformations along the long-axis where fibers are under tension compared to the short-axis where fibers are predominantly in compression (Figure 2). The inset (A) shows the deformation of the fibers and the inset (B) shows the deformation field from experiment[15]. The major mechanisms of deformation in networks of random fibers subjected to loading states such as simple shear and uniaxial tension are axial stretching of fibers along the principal tensile direction, bending of fibers under transverse loading and buckling of fibers under compression. Despite the complexity of spatial heterogeneity in loading resulting from the contraction of a single cell, we find that the same mechanisms are operative. The response at large strains is predominantly dictated by the constitutive response of the aligned fibers undergoing axial stretching [28]. The bending resistance of fibers is low compared to stretching and under applied tractions, fibers easily bend and reorient along the principal loading direction. The initial phase of deformation below a critical strain is dominated by fiber bending and is short ranged [19]. With further strain, fibers transition to stretching, with this transition identified by computing the strain at which the bending energy becomes equal to the stretching energy (see Appendix B for the energy transition in a single fiber and a network). The ratio of bending energy to stretching energy has been used to identify regimes with tension-dependent alignment of fibers [24], [29]. We adopted a similar strategy to define a *zone of the influence* around a cell up to which the cell contraction is felt. When the stretching energy of a fiber exceeds the bending energy, we assume that it is within the zone of influence; this criterion is computed using energy ratio $E_R = E_{stretch}/E_{bend} > 1$.

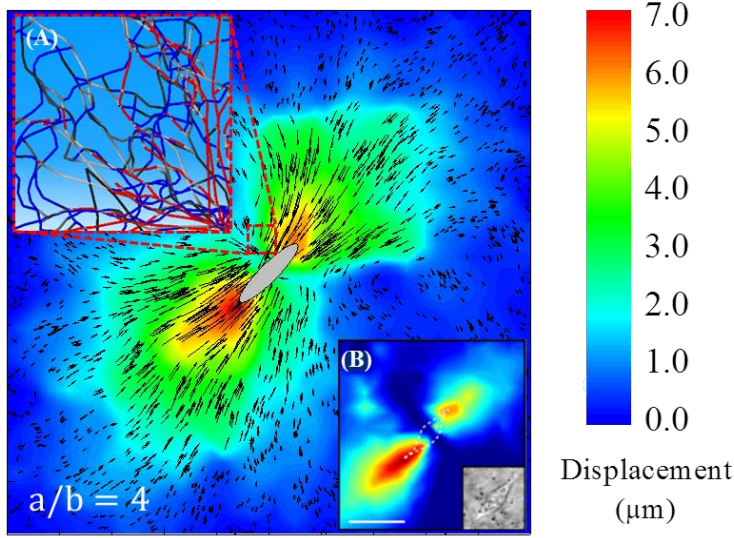

Figure 2: Displacement field after contraction of the cell (strain ~ 90%). The ellipse (a/b=4) at the center schematically shows the position of the cell before contraction. Inset A shows the aligned (red) and buckled (blue) fibers near the cell. Deformation is highly heterogeneous and localizes along the major axis of the cell as observed in experiments (inset B, adapted with permission from Gjorevski et al[15]). Displacement vectors show the magnitude and direction of the deformation. Note the longer aligned vectors along the long axis and shorter vectors along the short axis of the cell and their random orientations.

### 3. 1 The role of cell aspect ratio on deformation localization and collagen reorganization

The aspect ratio of the cell is linked to intrinsic cell behavior (for example, invasive vs. non-invasive tumor cells) and the influence of the local physical and chemical microenvironment. The shape of the cell also affects the polarization and collagen realignment[11]. In order to understand the role of cell shape, we considered cells of varying aspect ratios, a/b=1, 4 and 16 as shown in in Figure 3 a-c. The cell shape anisotropy leads to deformation anisotropy with large deformations along the longer axis (LA) of the cell. As the cell aspect ratios increase (at fixed cell area of 50 $\mu m^2$), larger strains are observed along the LA producing a localization of high strain energy density (SED) along the poles, while buckled fibers are observed along the shorter axis (SA) of the cell. Since buckling primarily involves bending of filaments, which is a soft mode of deformation, the SED in this case is small.

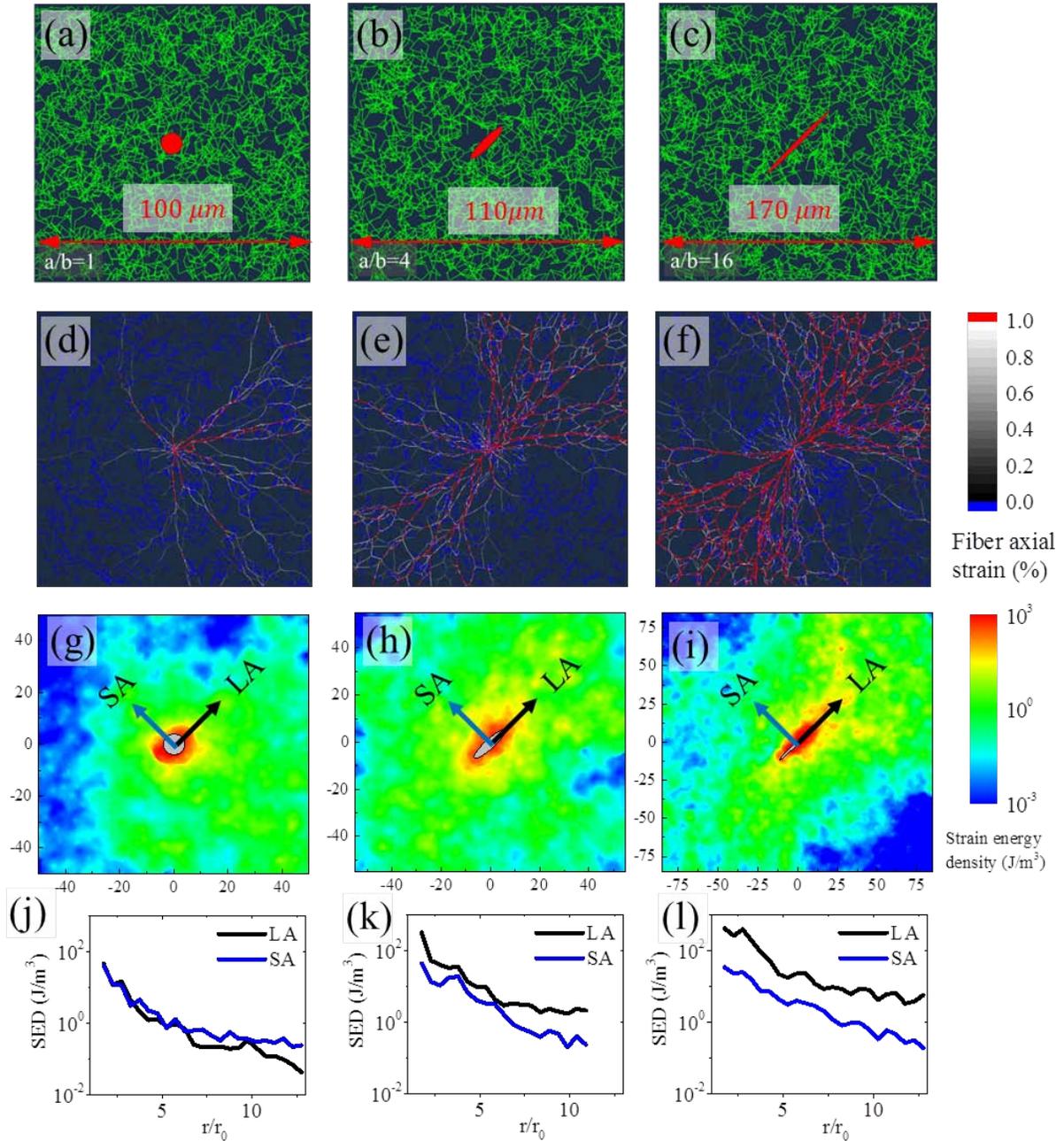

Figure 3: Role of cell aspect ratio on fiber reorganization and deformation anisotropy. (a-c) Shows the networks with cells (aspect ratios a/b=1, 4 and 16). (d-f) Fibers along the longer cell axis experience large axial stretching and preferential alignment at higher cell aspect ratios. Axial strains of red fibers exceed 1%, blue fibers are in compression and black to white denote fibers with strains in the range 0-1%. (g-i) Strain energy density (SED) of the networks at 25% cell contraction. (j-l) The averaged SED variation is similar along both the LA and the SA for the circular cell and localizes along the LA as the cells become more elliptical.

SED distribution is similar along both the SA and the LA of the circular cell (Figure 3 g) whereas for the elliptical cells (Figure 3 h-i), SED along the LA is greater than the SA. This deformation anisotropy is further quantified by computing the averaged variation of SED as a function of distance from the surface of the cell (Figure 3 j-l). For a circular cell, average SED variation is similar along both LA and SA and have a value of ~ $40 \, J/m^3$ at the cell surface and quickly decays with distance[1]. At an aspect ratio of a/b=4, SED along the LA (~ $300 \, J/m^3$) is larger than along the SA (~ $40 \, J/m^3$) due to the localized deformation along the poles. A similar albeit exaggerated trend is observed with further increases in the cell aspect ratio: when a/b=16, the energy offset is more than an order of magnitude (LA~ $400 \, J/m^3$ and SA~ $30 \, J/m^3$). The variation of SED observed in our simulation is qualitatively similar to the experimentally observed variations (SED~ $10 \, J/m^3$ [11]).

In the case of collagen gels when the local deformation exceeds the critical strain (called the knee strain[2]), fibers deform predominantly by axial stretching. In our simulations, we have observed that the fibers that align with the LA get stretched excessively (Figure 3 d-f). We have shown fibers with axial strains exceeding 1% in red and fibers under compression in blue. These images clearly show the increase in the fraction of aligned fibers along LA as the cell shape anisotropy increases. During the uniform contraction of a spindle shaped cell, shortening of the LA is much more than that of the SA, inducing compression in the material alongside LA. Hence the fibers along the LA are stretched (red fibers) while the fibers along the SA are compressed (blue fibers). This also explains the localization of SED along the poles (Figure 3 g-i). Taken together, these simulations highlight how in fibrous networks, fibers tend to rotate and re-orient towards the principal deformation axes as the cells contract and this effect is influenced by the cell aspect ratio, where at higher aspect ratios, there is more alignment along the LA than the SA (refer to Appendix C for plots of fiber orientation and its quantification).

**3.2 The role of cell contractility and aspect ratio on the range of force transmission**

Next, we focus on the role of cell contractility on the anisotropy of deformation fields. We find that with increasing contractility, more fibers are aligned, leading to the formation of collagen tracts (Figure 3 d-f). To study the role of cell contractility and aspect ratio in force transmission, the ratio of fiber stretching and bending energies ($E_R$) for circular (a/b=1) and spindle shaped (a/b=16) cells at various levels of contraction are plotted as a function of distance from the cell surface (Figure 4 a-b). The plot has two characteristic zones, a bending dominated zone ($E_R<1$) and a stretching dominated zone ($E_R>1$) shown by the shaded region in the plot. The energy ratio at a distance $r$ from the cell center is the averaged value of all elements in a concentric circular region at a distance $r \pm 1 \, \mu m$ from the center.

---

[1] The nominal strain energy density of a square sample of the fibrous material in simple shear at a strain of 10% is~ $50 \, J/m^3$ .
[2] The characteristic strain at which the gels stiffens. For collagen in tensile loading, its ~ 10% [39].

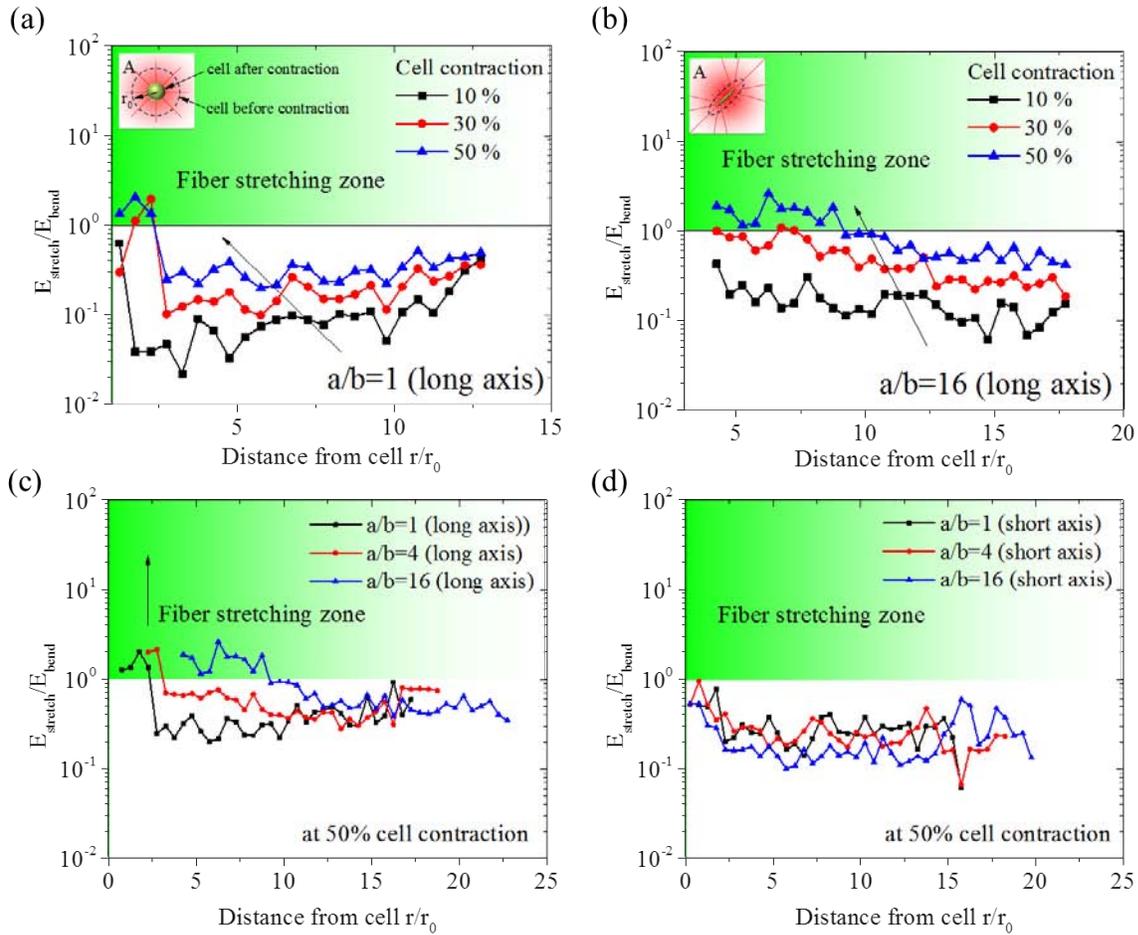

Figure 4: Plot of the ratios of stretching to bending energies ($E_R = E_{stretch}/E_{bend}$) as a function of the distance from the cell surface for different levels of contractility and aspect ratios. The shaded zone corresponds to the stretching dominated regime. At higher levels of cell contractility and aspect ratios, more fibers are axially stretched and the deformation becomes stretching dominated. (a) For the circular cell at 40-50% cell contractility the size of the fiber stretching dominates region is only $r = 2.5\ r_0$. (b) When the cell aspect ratio increases, at the same levels of contractility, the deformation zone extends up to $r = 10 r_0$. Inset in both figures schematically shows the cell shapes before and after contraction. (c) $E_R$ variation along LA at 50% strain for three aspect ratios. Cells of all aspect ratios have a stretching dominated zone up to certain distance from the cell surface and this increases with the aspect ratio. (d) $E_R$ along SA– in this case the deformation is entirely bending dominated.

Elements near the cell surface transition from being bending dominated to stretching dominated as cell contractility increases. The stretching dominated zone extends further away from the cell surface with increasing aspect ratios. For a/b=1 at 50% contractility, the deformation zone extends up to a distance of $r \sim 2.5\ r_0$, where $r$ is the radial distance from the cell and $r_0$ is the initial cell radius of the circular cell (see Figure 4 b inset (A) for the definition of $r_0$). As the cell becomes more spindle shaped, the deformation zone extends further away and at the largest ratio we consider, a/b=16, $r \sim 10\ r_0$. Variation of $E_R$ at 50% cell contractility for three aspect ratios as a function of the distance from the cell shows the dominant role played by the aspect ratio in the range of force transmission and deformation anisotropy (Figure 4 c-d). The size of deformation zone around a cell for the same level of contractility is bigger at larger aspect ratios. For all aspect ratios, deformation localizes along the LA and the stress fields are transmitted to longer distances compared to the SA.

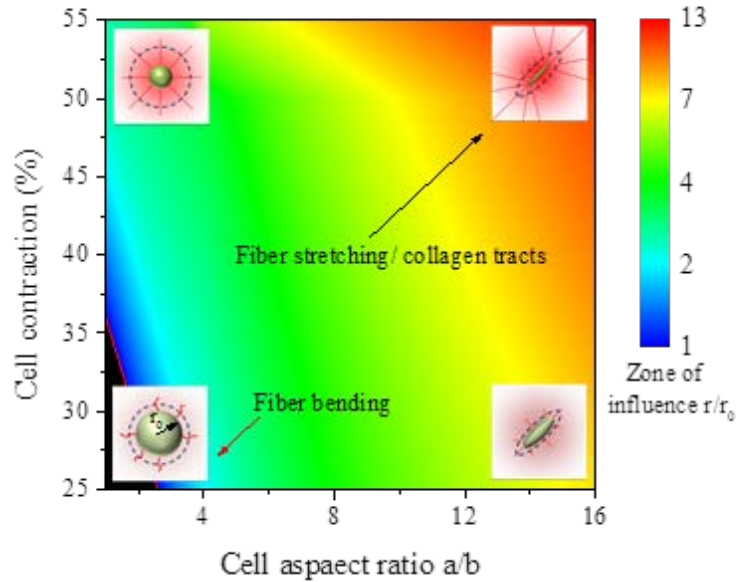

Figure 5: Heat map of the deformation zone around a single contracting cell. The zone of influence increases with both contractility and the shape anisotropy. For a circular cell, up to a contractile strain of 35%, the gel around the cell is in the bending-dominated regime of deformation and fiber alignment is not observed. As the contractility increases, collagen fibers are aligned akin to collagen tracts seen in experiments (top left corner). Note that a very strong alignment is seen for spindle shaped cells at large contractility (top right corner).

To summarize these findings, a heat map of the zone of influence based on the energy criterion, $E_R > 1$ around a single cell for different cell aspect ratios and contractility is developed (Figure 5). This map predicts the size of the deformation zone for cells of shapes ranging from circular to highly spindle shaped and for contractile strains as high as 55%. The black region at the lower left corner corresponds to $E_R < 1$ (for smaller aspect ratios and lower contractile strains). When the cell is circular, for contractile strains up to ~35%, the deformation is bending dominated whereas at higher cell aspect ratios, anisotropic deformation around the poles is enough to axially stretch the fibers and aligned bundles of collagen fibers form around the cell as observed in Figure 3 (g-h). Insets in Figure 5 show the approximate cell shape and the nature of the collagen fibers around the cell in the respective regions of the map, where dotted tracts show the initial cell size and the colored solid shapes show the shape of the cell after contraction.

## 3.3 Strong cell-cell interaction by the formation of collagen tracts at smaller separation distances and larger aspect ratios

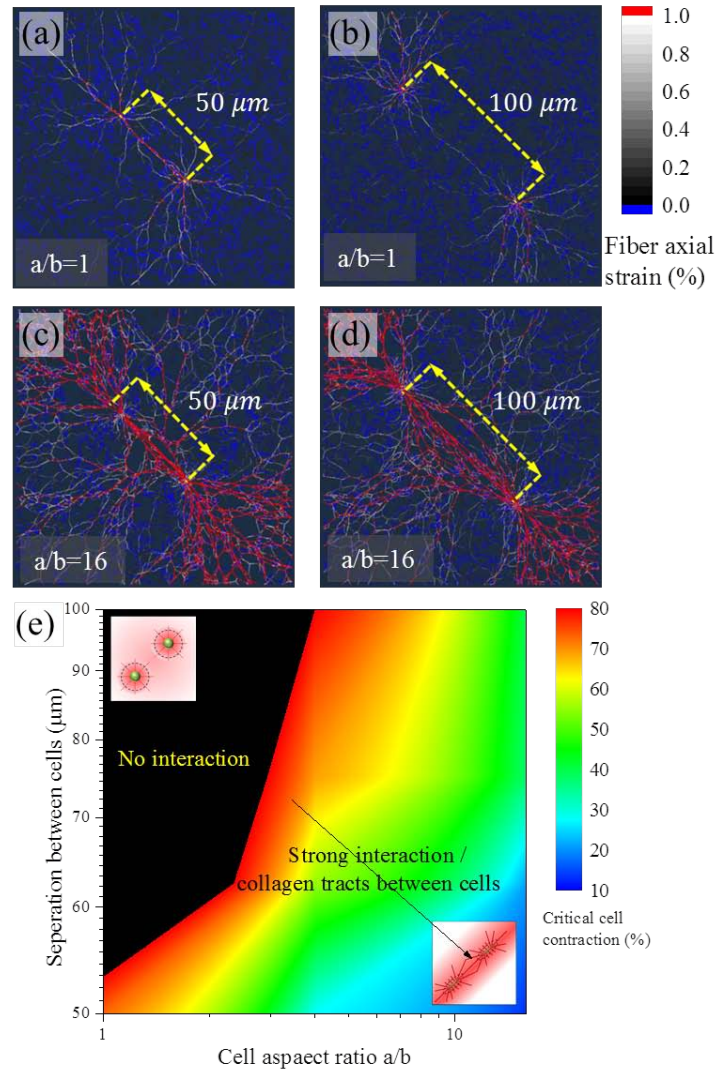

Figure 6: Interaction between two cells with varying center to center distance at ~90% cell contraction. When the distance is 50 $\mu m$, cells of all aspects ratios interact by forming collagen tracts. However, as the separation distance increases, the collagen tracts in the deformation zones around the circular cells do not interact with each other. Axial strains of red fibers exceed 1%, blue fibers are in compression and black to white denote fibers with strains in the range 0-1%. (e) Heat maps of critical contraction strains for interaction between two cells as a function of their separation. The interaction is strong at higher aspect ratios and smaller separation distances. For circular cells, when the separation between cells is more than 50 $\mu m$, no visible collagen tracts are observed.

In a historical experiment demonstrating how cellular forces actively remodel the surrounding collagen matrix, Stopak and Harris cultured tissue explants and observed the formation of aligned tracts of collagen fibers between them [30]. As increasing contractility is one way in which cells and tissues respond to external forces, these experiments demonstrated how multicellular mechanical signaling could be transduced through alignment of the ECM. The remodeling of collagen is key to tumor cell migration, where aligned tracts of fibers enable directionally persistent migration [31]. In the case of collective migration, cells follow the pre-existing space and the aligned migration tracks [32]. As a first step towards understanding the collagen reorganization in cell-seeded matrices, we model the response of two cells of varying aspect ratios and separation. The energy ratio ($E_R$) is used to identify the nature of the deformation zone around the cell and the critical cell contractility at which two cells start to interact.

As the anisotropy and contractility increases, more collagen fibers around the cell deform by axial stretching leading to numerous fibers extending radially from the cell (Figure 6). When the deformation zones of the two cells overlap, highly aligned and stretched fibers form between them (Figure 6 a-b). When the separation distance is 50 $\mu m$, cells form weak collagen tracts. At a separation distance of 100 $\mu m$, although collagen tracts radially emanate from the cells, they do not interact and no tracts are established connecting the two cells (Figure 6 b). There is a significant difference in the interaction pattern of cells as the aspect ratio changes (Figure 6 c-d). At the given contractility of ~90%, cells interact strongly through the formation of aligned and stretched fibers. When the separation distance is smaller, the interaction is stronger and thick bundles of aligned fibers are formed. Thus, both cell aspect ratio and cell spacing factor into long distance intercellular communication.

We systematically investigate the role of cell aspect ratio and separation distance in interaction through a parametric study [3]. The separation distance is varied in increments of 12.5 $\mu m$ and the cell contractility at which collagen tracts are established is estimated using the magnitude of $E_R$. Results of the parametric study are summarized in a heat map in Figure 6 e. The black zone at lower aspect ratios corresponds to zone where cells do not interact. Even at very large contractile strains, the induced deformation is not sufficient to cause a transition of the response of the fibers from the bending dominated to the stretching dominated regime. With increasing aspect ratios, the influence zone extends anisotropically along the LA and interferes with the influence zone of the nearby cell. At 50 $\mu m$ separation, for a/b=16, the critical contractile strain is only~10%. At smaller cell aspect ratios of a/b~1, visible collagen tracts are formed only at a contractile strain of~80%.

### 3.4 DFN model predicts longer range of force transmission compared to the Neo-Hookean material model

---

[3] In all simulations, the LAs of the cells are aligned on a straight line with varying separation distance. Offset between the LA and different cell orientations are not considered in the current work.

Soft-materials like fibrous matrices show non-linear hardening response and sustain very large strains. This is akin to the response of synthetic materials such as polymer gels and rubbers. One of the most commonly used material models to describe such a response is the neo-Hookean constitutive law. However, in recent experiments by the group of Billiar and Hart, it has been observed that the range of force transmission in collagen gels is much higher than that observed in synthetic materials like polyacrylamide gels and they attributed this to the presence of fibers [3], [4]. They further verified this concept by computational models using conventional (NH) material models. To further understand the differences between isotropic strain hardening materials (with the neo-Hookean model as an example) and fibrous materials, we carried out a comparative study of our discrete fiber network model with a similar finite element model with the neo-Hookean (NH) material.

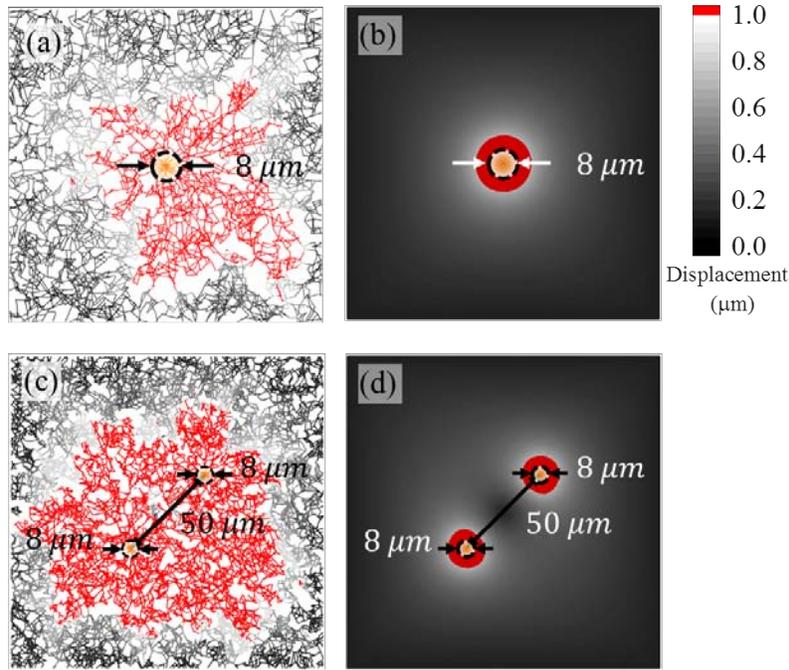

Figure 7: Displacement fields of the discrete fiber network model and the neo-Hookean material. The cell is circular and the model size is identical in both cases. (a) The displacement is felt at a radius of $r \sim 4r_0$ for the fibrous network. (b) For the equivalent 2D plane stress model with a non- linear hardening neo- Hookean (NH) material, the field in red is only significant for $r \sim r_0$. (c) Displacement fields of two cells separated by $50 \mu m$ overlap with each other for the discrete fiber network model. (d) For the same separation distance and contractility, the displacement fields of NH material remain isolated and no interaction is observed. Regions with displacements exceeding $1 \mu m$ are shown in red and the initial cell size is schematically shown by a circle ($r_0 = 4 \mu m$).

The response of the cell contracting in fibrous networks modeled using the discrete fiber network approach (a) and the equivalent 2D plane stress model with neo-Hookean material is shown in Figure 7. To study the extent of influence, material displacements exceeding $1 \mu m$ are shown in red. It is clear that the DFN model shows a very large red zone around the cell while the NH model shows only a very small zone. The displacement fields for DFN model extends up to 4 times the initial cell size ($r \sim 4r_0$) while the displacement zone for the non-linear hardening NH model decays rapidly with the distance and is only in the range, $r \sim r_0$ (Figure 7 a-b). This clearly shows that the material non-linearity alone is not sufficient to explain the long range force transmission observed in fibrous matrices. We also carried out simulations to study the interactions of two cells (Figure 7 c-d) and found that in the case of the NH model, the displacement zones of the two cells do not overlap, which implies that they do not interact. On the other hand, for the same set of material and geometric parameters, the cells in the fibrous matrix interact strongly.

The ability of the DFN model to predict the response of cell-seeded matrices for single and two cells is evident from these simulations. Comparison with NH materials shows the inability of conventional strain hardening material models to predict the large deformation zones of contracting cells. When the aspect ratio of the cell increases, the deformation zone extends further along the LA. This results in stronger interaction between cells. The NH model also shows a small increase in the deformation along the LA for higher aspect ratios, but is not significant compared to the DFN model.

## 4. Summary

We have developed a FE based discrete fiber model to simulate the response of cell populated fibrous matrices. The initial fiber distribution of the matrices is assumed to be random and the model naturally captures fiber reorganization, alignment and bundling observed in experiments. The results of the parametric study for variations in cell shape and contractility are summarized in a phase diagram to make testable predictions on the zone of influence around a contracting cell in fibrous matrices. The interactions between two cells are studied using a similar approach and the critical contractility for the cell-cell interaction as a function of cell aspect ratio and separation distance is also described using a phase diagram. The primary predictions from the current work are

1. Long range force transmission: Our results clearly show that the long-range displacement fields within matrices can be captured by tension-driven local fiber alignment along principal directions of stretch, and that the heterogeneities result from the anisotropic shape of the cell domain. We started with a uniform fiber distribution of the fibrous material and the model predicted the reorganization, alignment and stretching of fibers akin to experiments [16], [34].
2. Cell shape anisotropy: The role of cell shape anisotropy on the range of force transmission and cell-cell interaction is systematically studied by varying the aspect ratio of the cell from circular to long spindle shapes. The circular cell has a uniform deformation zone around it while large tensile deformation zone is identified along the poles of the elongated spindle shaped cells. Fibers along the short axes of elongated cells are found to be in compression as observed in experiments. Forces are transmitted for a longer distance along the long cell axis compared to the shorter axis in the case of elongated cells.

3. Energy based criterion and deformation zones: The characteristics *knee strain* associated with the response of collagen gels is due to the cross over from bending to stretching modes of deformation with increasing strain. We defined an energy based parameter $E_R = E_{stretch}/E_{bend}$ by computing the stretching and bending energy of deforming fibers and identified the fibers around the cell which transition to stretching dominated regime based on $E_R > 1$. We have identified a zone of influence around a contracting cell based on the ratio of stretching to bending energy, $E_R$. Using this approach, we carried out a parametric study over a range of cell aspect ratios and contractile strains and developed a phase diagram to predict the response of cell-seeded matrices.
4. Interaction between cells: The critical cell contractility required for the interaction between two cells by the formation of aligned collagen fibers is identified. As in the case of single cells, we derived a phase diagram for cell-cell interactions as a function of their separation distance and aspect ratios.
5. Comparison with isotropic strain hardening models: We have shown that the long range force transmission observed in collagen gel is due to the fibrous nature of the matrices and conventional strain hardening material models are incapable of predicting this. We compared our DFN model with an equivalent Neo-Hookean model and observed that the range of force transmission in DFN models is several times of that of the NH model.

*Comparison to experiments:* Gjorevski and Nelson examined the strain fields around engineered epithelial tissues in collagen I gels. They found that linear elasticity cannot explain the long-range nature of the strain fields but reported that mechanical heterogeneities caused by stiffening near the poles of elongated contractile epithelial tissues can explain the decay of strain fields [15]. Contraction of cells can reorganize the ECM to provide contact guidance that facilitates migration/ invasion [2], [8], [33]. These fiber alignments observed between nearby cells in matrices [2], [8], [33] are clearly shown in our FEM simulations (Figure 2 and 3). Treatment of the cells to abolish actomyosin contractility leads to dissolution of the collagen lines, in agreement with our results that show that the magnitude of contractile strains play an important role in determining the range of force transmission as shown in Figure 4. In contrast to cells on PA gels, human mesenchymal stem cells (hMSCs) and 3T3 fibroblasts on fibrin gels were shown to sense and respond to mechanical signals up to five cell lengths away [34], consistent with our results shown Figures 3 and 4. Furthermore, Ma et al. suggest that the fibrous nature of the ECM leads to reorganization of the collagen fibers leading to areas of higher fiber density near the cells over relatively long distances (10 cell-diameters) [35]. The mechanism whereby this reorganization proceeds (starting from a random network) is discussed in our work. Finally, Koch et al. studied the effect of anisotropic cell shape and contractility on range of force transmission in invasive and non-invasive cancer cells [36]. They found that both lung and breast carcinoma cells were significantly elongated compared to the non-invasive cells, which were observed to have rounder shapes. Cell shape anisotropy was accompanied by a larger sensing distance and stiffening and fiber at the poles, suggesting that directionality of traction forces is important for cancer cell invasion, consistent with our results (Figures 2 and 3).

**Appendix A: Conversion of collagen concentration to equivalent network density**

Collagen gel considered in experiments is converted into a computational network (with equivalent fiber density) using the approach of Stein, Andrew M., et al [37]. For the given concentration and volume of the gel, fiber radius is given by

$$r = \sqrt{\frac{V_g \rho_c v_c}{\pi L_{Tot}}}$$

where $V_g$ $(\mu m^3)$ is the volume of the gel, $\rho_c (= 1 - 5 \ mg/ml)$ is the mass density of collagen, $v_c = 0.73 \ ml/g$ is the specific volume of collagen, $r$ $(\mu m)$ is the radius of the fibers and $L_{Tot}$ $(\mu m)$ is the total length of collagen in the gel. The 3D variables are converted into equivalent 2D ones by transforming quantities per unit volume to quantities per unit area. Fiber radius is assumed to be $250 \ nm$ and from the above relation, the total length of fiber in the gel is calculated for varying collagen concentrations. The fibers have both flexural and stretching rigidities and the crosslinks are assumed to be rigid[22]. A parametric study for various collagen concentrations $(2, 3, 4 \ and \ 5 \ mg/ml)$, simulating simple shear deformation (Figure 8) shows good agreement with the experimentally observed strain sweep results [38]. Increasing gel concentration reduces the collagen mesh size (distance between two crosslinks) leading to a stiffer response. The reduction in the length of the fiber between the crosslinks affects the bending characteristics and leads to an increase in the initial stiffness and a decrease in the knee strain.

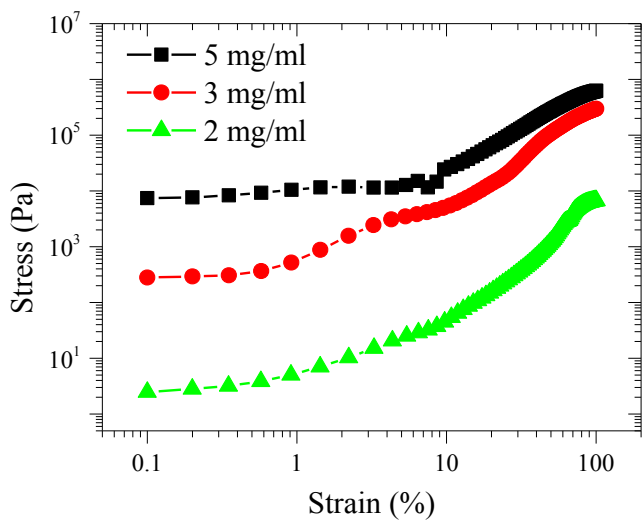

Figure 8: Simple shear deformation of collagen network with varying gel concentrations. The characteristic non-linear strain hardening response with the concentration dependent stiffening observed in experiments is captured well by the model.

The model shows the characteristic response of collagen gels; non-linear strain hardening and concentration dependent stiffening. Having developed the basic framework for fibrous materials, we next tried to understand the basic mechanisms and energy of different deformation modes associated with this response.

**Appendix B: Energy of fibers during network deformation**

Collagen gels show a characteristic non-linear strain hardening response with a "knee" around 10% strain. This is typical with most bio-materials and the onset of hardening is associated with the bending-stretching transition of the fibers that constitute the gels. We verified this concept both at a single fiber and network (gel) levels (Figure 9). The non-linear response can be approximated by a bi-linear response with a change in slope at the knee. It is schematically shown in the figure by black dashed lines. The initial compliant response is bending dominated and at the knee, there is cross over to a regime where stretching energy starts to dominate the overall response. For a single fiber, the transition point is well defined (Figure 9 a) whereas for the network, this takes place over a range of strains around 10-15% (Figure 9 b).In the case of the network, individual fibers at various locations of the gel get strained at different levels, and at this point the overall energy becomes stretching dominated.

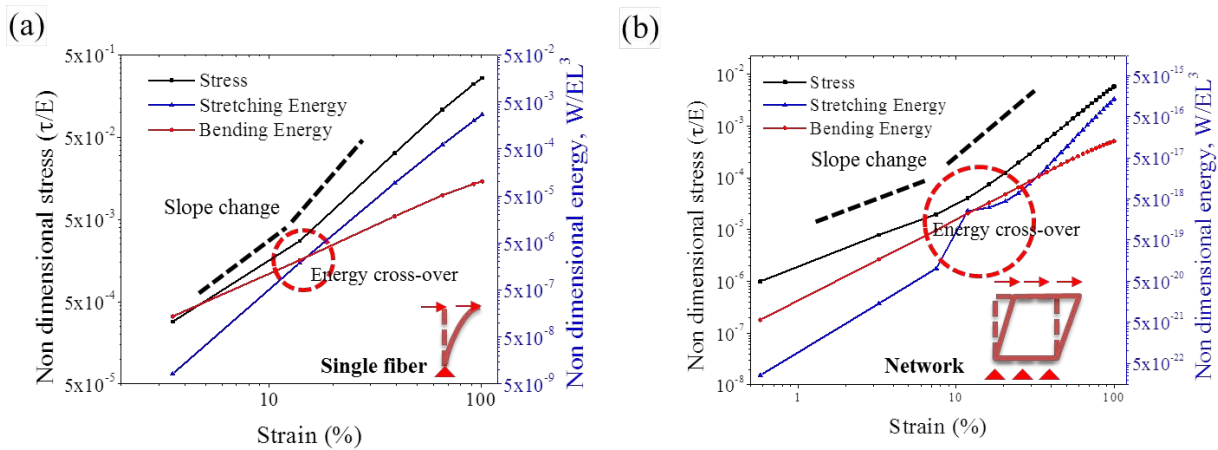

Figure 9: Stress- strain response and energy of deformation. (a) Response of a single fiber fixed at the bottom and deformed at the top as shown in the inset. The initial deformation is dominated by fiber bending and at ~$\mathbf{10\%}$ strain, stretching energy dominates the response. (b) A similar response is observed for networks, however, instead of a well-defined knee, it occurs over a range of strains. In the case of a network, individual fibers make transitions as shown in (a) and the cooperative response is shown in (b).

Having gained the understanding of the energy transition both at single fiber and network level, we used this approach to define the deformation zone for the response of matrices with contractile cells. We computed the energy ratio, $E_R = E_{stretch}/ E_{bend}$ of all elements in the computational domain. At the knee, $E_R = 1$, fibers transition to stretching dominated response.

**Appendix C: Quantification of fiber alignment**

During cell contraction, there is large scale reorientation of fibers. We identified the long and the short axes of the cell as shown in Figure 10. The quadrants constitute regions $\pm 45^0$ on the two sides of the short (SA) and long (LA) axes. In the quadrant marked LA in Figure 10, fibers get stretched well past 1% strain which is evident from large number of red fibers which align along the LA of the cell. Nearly 5% of the fibers align along LA and these aligning fibers are recruited from the randomly oriented fibers (Figure 10 b). There is a decrease in the fibers oriented at $90^0 - 180^0$ and these reorient towards the LA. Fraction of fibers aligning to SA is less than 2.5% (Figure 10 b).

We further quantified the alignment by computing a dimensionless orientation parameter. For the 2D case, it is defined as $S = <2\cos^2\theta - 1>$. Fiber alignment during the cell contraction is quantified by computing the difference in $S$ before and after contraction as $\Delta S = S_{final} - S_{initial}$. The anisotropy in alignment increases as cells becomes elliptical. $\Delta S$ for a circular cell (Figure 11 a) is similar along both axes whereas for a spindle shaped cell, $\Delta S$ is only 4% along SA while it is 10% along LA at very large contractility (Figure 11 b).

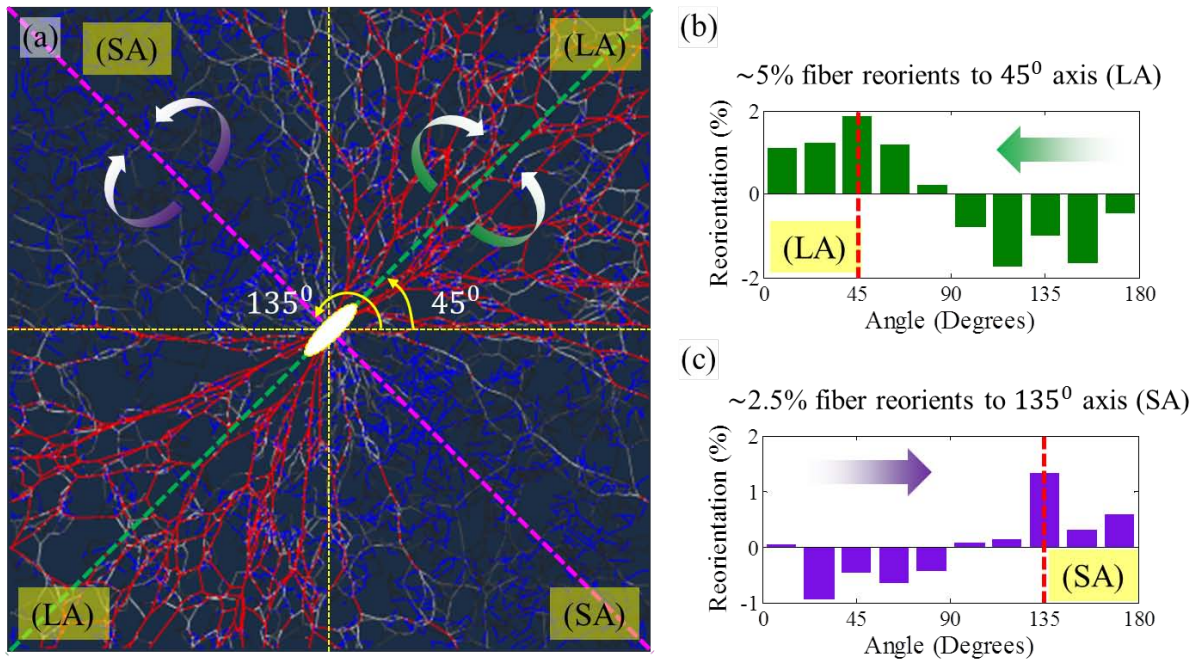

Figure 10: Contraction induced fiber reorientation and alignment. Quadrants encompassing long axes of the cell at $45^0$ and short cell axis at $135^0$ are labeled as LA and SA respectively. During the contraction of cell, fibers align to the axes and the fraction of fibers aligning to LA is shown in (b) and SA in (c).

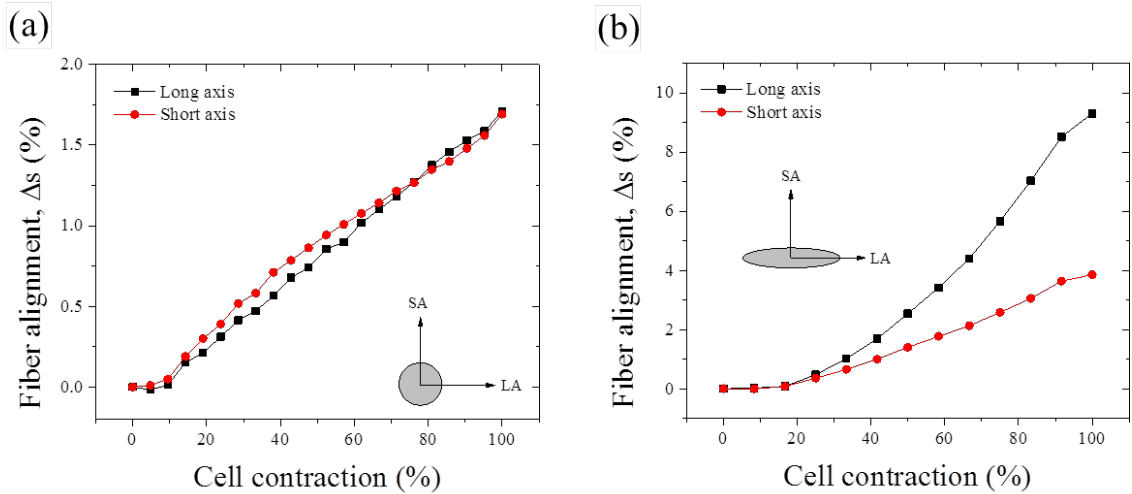

Figure 11: Fiber alignment computed as the change in orientation parameter before and after cell contraction. Alignment is a function of cell aspect ratio and contractility. More fibers align to the axis of contraction and as the aspect ratio increases and there is a pronounced alignment along longer cell axis. (a) When the cell is circular, alignment along LA and SA is similar. (b) For the spindle shaped cell, alignment along LA is more than along SA.

## Acknowledgements

Research reported in this publication was supported by the National Institute of Biomedical Imaging and Bioengineering of the National Institutes of Health under Award Number R01EB017753 and the US National Science Foundation Grant CMMI-1312392. The content is solely the responsibility of the authors and does not necessarily represent the official views of the National Institutes of Health or the National Science Foundation.

## References

[1] H. Y. Chang, J. B. Sneddon, A. a Alizadeh, R. Sood, R. B. West, K. Montgomery, J.-T. Chi, M. van de Rijn, D. Botstein, and P. O. Brown, "Gene expression signature of fibroblast serum response predicts human cancer progression: similarities between tumors and wounds.," *PLoS Biol.*, vol. 2, no. 2, p. E7, Feb. 2004.

[2] A. K. Harris, D. Stopak, and P. Wild, "Fibroblast traction as a mechanism for collagen morphogenesis.," *Nature*, vol. 290, pp. 249–251, 1981.

[3] X. Ma, M. E. Schickel, M. D. Stevenson, A. L. Sarang-Sieminski, K. J. Gooch, S. N. Ghadiali, and R. T. Hart, "Fibers in the extracellular matrix enable long-range stress transmission between cells," *Biophys. J.*, vol. 104, no. 7, pp. 1410–1418, Apr. 2013.

[4] M. S. Rudnicki, H. a Cirka, M. Aghvami, E. a Sander, Q. Wen, and K. L. Billiar, "Nonlinear strain stiffening is not sufficient to explain how far cells can feel on fibrous protein gels.," *Biophys. J.*, vol. 105, no. 1, pp. 11–20, Jul. 2013.

[5]     J. P. Winer, S. Oake, and P. a Janmey, "Non-linear elasticity of extracellular matrices enables contractile cells to communicate local position and orientation.," *PLoS One*, vol. 4, no. 7, p. e6382, Jan. 2009.

[6]     P. P. Provenzano, K. W. Eliceiri, J. M. Campbell, D. R. Inman, J. G. White, and P. J. Keely, "Collagen reorganization at the tumor-stromal interface facilitates local invasion.," *BMC Med.*, vol. 4, no. 1, p. 38, Jan. 2006.

[7]     M. Vishwanath, "Modulation of Corneal Fibroblast Contractility within Fibrillar Collagen Matrices," *Invest. Ophthalmol. Vis. Sci.*, vol. 44, no. 11, pp. 4724–4735, Nov. 2003.

[8]     P. P. Provenzano, D. R. Inman, K. W. Eliceiri, S. M. Trier, and P. J. Keely, "Contact guidance mediated three-dimensional cell migration is regulated by Rho/ROCK-dependent matrix reorganization.," *Biophys. J.*, vol. 95, no. 11, pp. 5374–5384, Dec. 2008.

[9]     M. W. Conklin, J. C. Eickhoff, K. M. Riching, C. A. Pehlke, K. W. Eliceiri, P. P. Provenzano, A. Friedl, and P. J. Keely, "Aligned collagen is a prognostic signature for survival in human breast carcinoma," *Am. J. Pathol.*, vol. 178, pp. 1221–1232, 2011.

[10]    P. P. Provenzano, K. W. Eliceiri, J. M. Campbell, D. R. Inman, J. G. White, and P. J. Keely, "Collagen reorganization at the tumor-stromal interface facilitates local invasion.," *BMC Med.*, vol. 4, no. 1, p. 38, Jan. 2006.

[11]    T. M. Koch, S. Münster, N. Bonakdar, J. P. Butler, and B. Fabry, "3D traction forces in cancer cell invasion," *PLoS One*, vol. 7, no. 3, 2012.

[12]    L. M. Sander, "Alignment localization in nonlinear biological media.," *J. Biomech. Eng.*, vol. 135, no. 7, p. 71006, Jul. 2013.

[13]    M. Aghvami, V. H. Barocas, and E. a Sander, "Multiscale mechanical simulations of cell compacted collagen gels.," *J. Biomech. Eng.*, vol. 135, no. 7, p. 71004, Jul. 2013.

[14]    X. Ma, M. E. Schickel, M. D. Stevenson, A. L. Sarang-Sieminski, K. J. Gooch, S. N. Ghadiali, and R. T. Hart, "Fibers in the extracellular matrix enable long-range stress transmission between cells," *Biophys. J.*, vol. 104, no. 7, pp. 1410–1418, Apr. 2013.

[15]    N. Gjorevski and C. M. Nelson, "Mapping of mechanical strains and stresses around quiescent engineered three-dimensional epithelial tissues," *Biophys. J.*, vol. 103, no. 1, pp. 152–162, Jul. 2012.

[16]    E. a Sander, T. Stylianopoulos, R. T. Tranquillo, and V. H. Barocas, "Image-based multiscale modeling predicts tissue-level and network-level fiber reorganization in stretched cell-compacted collagen gels.," *Proc. Natl. Acad. Sci. U. S. A.*, vol. 106, no. 42, pp. 17675–17680, 2009.

[17]    M. F. Hadi, E. a Sander, J. W. Ruberti, and V. H. Barocas, "Simulated remodeling of loaded collagen networks via strain-dependent enzymatic degradation and constant-rate fiber growth.," *Mech. Mater.*, vol. 44, pp. 72–82, Jan. 2012.

[18]  M. C. Evans and V. H. Barocas, "The modulus of fibroblast-populated collagen gels is not determined by final collagen and cell concentration: Experiments and an inclusion-based model.," *J. Biomech. Eng.*, vol. 131, no. 10, p. 101014, Oct. 2009.

[19]  P. Onck, T. Koeman, T. van Dillen, and E. van der Giessen, "Alternative Explanation of Stiffening in Cross-Linked Semiflexible Networks," *Phys. Rev. Lett.*, vol. 95, no. 17, p. 178102, Oct. 2005.

[20]  B. Hu, V. B. Shenoy, and Y. Lin, "Buckling and enforced stretching of bio-filaments," *J. Mech. Phys. Solids*, vol. 60, no. 11, pp. 1941–1951, Nov. 2012.

[21]  P. Chen and V. B. Shenoy, "Strain stiffening induced by molecular motors in active crosslinked biopolymer networks," *Soft Matter*, vol. 7, no. 2, pp. 355–358, 2011.

[22]  A. Nair, "Discrete Micromechanics of Random Fibrous Architectures," PhD thesis, National University of Singapore, 2012.

[23]  A. Nair, "Discrete Micromechanics of Random Fibrous Architectures," *PhD Thesis, Natl. Univ. Singapore*, vol. Chapter 2, 2012.

[24]  a. Shahsavari and R. C. Picu, "Model selection for athermal cross-linked fiber networks," *Phys. Rev. E*, vol. 86, no. 1, p. 011923, Jul. 2012.

[25]  E. Gentleman, A. N. Lay, D. A. Dickerson, E. A. Nauman, G. A. Livesay, and K. C. Dee, "Mechanical characterization of collagen fibers and scaffolds for tissue engineering," *Biomaterials*, vol. 24, no. 21, pp. 3805–3813, 2003.

[26]  D. L. Christiansen, E. K. Huang, and F. H. Silver, "Assembly of type I collagen: fusion of fibril subunits and the influence of fibril diameter on mechanical properties," *Matrix Biol.*, vol. 19, no. 5, pp. 409–420, 2000.

[27]  Hibbitt, Karlsson, and Sorensen, *ABAQUS/Standard user's manual*, vol. 1. Hibbitt, Karlsson & Sorensen, 2001.

[28]  Q. Wen, A. Basu, J. P. Winer, A. Yodh, and P. A Janmey, "Local and global deformations in a strain-stiffening fibrin gel," *New J. Phys.*, vol. 9, no. 11, pp. 428–428, Nov. 2007.

[29]  G. Žagar, P. R. Onck, and E. Van der Giessen, "Elasticity of Rigidly Cross-Linked Networks of Athermal Filaments," *Macromolecules*, vol. 44, no. 17, pp. 7026–7033, Sep. 2011.

[30]  D. Stopak and A. K. Harris, "Connective tissue morphogenesis by fibroblast traction," *Dev. Biol.*, vol. 90, no. 2, pp. 383–398, Apr. 1982.

[31]  P. Friedl, K. Maaser, C. E. Klein, B. Niggemann, G. Krohne, and K. S. Zã, "Migration of Highly Aggressive MV3 Melanoma Cells in 3-Dimensional Collagen Lattices Results in Local Matrix Reorganization and Shedding of α 2 and β 1 Integrins and CD44 Migration of Highly Aggressive MV3 Melanoma Cells in 3-Dimensional Collagen Lattices," pp. 2061–2070, 1997.

[32]  A. Haeger, M. Krause, K. Wolf, and P. Friedl, "Cell jamming: Collective invasion of mesenchymal tumor cells imposed by tissue confinement.," *Biochim. Biophys. Acta*, Apr. 2014.

[33] M. Miron-mendoza, J. Seemann, and F. Grinnell, "Collagen Fibril Flow and Tissue Translocation Coupled to Fibroblast Migration in 3D Collagen Matrices," vol. 19, no. May, pp. 2051–2058, 2008.

[34] J. P. Winer, S. Oake, and P. a Janmey, "Non-linear elasticity of extracellular matrices enables contractile cells to communicate local position and orientation.," *PLoS One*, vol. 4, no. 7, p. e6382, Jan. 2009.

[35] X. Ma, M. E. Schickel, M. D. Stevenson, A. L. Sarang-Sieminski, K. J. Gooch, S. N. Ghadiali, and R. T. Hart, "Fibers in the extracellular matrix enable long-range stress transmission between cells," *Biophys. J.*, vol. 104, no. 7, pp. 1410–1418, Apr. 2013.

[36] T. M. Koch, S. Münster, N. Bonakdar, J. P. Butler, and B. Fabry, "3D traction forces in cancer cell invasion," *PLoS One*, vol. 7, no. 3, 2012.

[37] A. M. Stein, D. a Vader, L. M. Jawerth, D. a Weitz, and L. M. Sander, "An algorithm for extracting the network geometry of three-dimensional collagen gels.," *J. Microsc.*, vol. 232, no. 3, pp. 463–75, Dec. 2008.

[38] A. M. Stein, D. A. Vader, D. A. Weitz, and L. M. Sander, "The Micromechanics of Three-Dimensional Collagen-I Gels," vol. 16, no. 4, 2011.

[39] B. a. Roeder, K. Kokini, J. E. Sturgis, J. P. Robinson, and S. L. Voytik-Harbin, "Tensile Mechanical Properties of Three-Dimensional Type I Collagen Extracellular Matrices With Varied Microstructure," *J. Biomech. Eng.*, vol. 124, no. 2, p. 214, 2002.